\documentclass[aps,prd,reprint,onecolumn,nofootinbib,letterpaper,preprintnumbers,
    superscriptaddress,eqsecnum]{revtex4}

\usepackage{longtable}
\usepackage{amsfonts}
\usepackage{amssymb}
\usepackage{amsmath}
\usepackage{amsthm}
\usepackage[usenames,dvipsnames]{color}
\usepackage{bm}
\usepackage{tensor}
\usepackage{graphicx}
\usepackage{hyperref}
\usepackage{dsfont}
\usepackage{array}
\usepackage{calc}
\usepackage{ifthen}
\usepackage{caption2}
\usepackage{setspace}

\hypersetup{
    bookmarks=true,
    unicode=false,
    pdftoolbar=true,
    pdfmenubar=true,
    pdffitwindow=false,
    pdfstartview={FitH},
    pdftitle={Systematic Investigations of the Free Fermionic Heterotic String
        Gauge Group Statistics: Layer 1 Results},
    pdfauthor={D. Moore, J. Greenwald, T. Renner, M. Robinson, and G. Cleaver},
    pdfsubject={Layer 1 FFHS Gauge Models},
    pdfcreator={D. Moore},
    pdfproducer={D. Moore},
    pdfkeywords={Gauge}{FFHS}{Surveys},
    pdfnewwindow=true,              % links in new window
    colorlinks=true,                % false: boxed links; true: colored links
    linkcolor=blue,                 % color of internal links
    citecolor=blue,                 % color of links to bibliography
    filecolor=blue,                 % color of file links
    urlcolor=blue                   % color of external links
}

\newcommand{\U}[1]{\ensuremath{U_{#1}}}
\newcommand{\SU}[1]{\ensuremath{SU_{#1}}}
\newcommand{\SO}[1]{\ensuremath{SO_{#1}}}
\newcommand{\E}[1]{\ensuremath{E_{#1}}}
\newcommand{\EE}[2]{\ensuremath{\E{{#1}}\otimes\E{{#2}}}}
\newcommand{\FSU}[1]{\ensuremath{\SU{{#1}}\otimes\U{1}}}
\newcommand{\PS}[0]{\ensuremath{\SU{4}\otimes\SU{2}\otimes\SU{2}}}
\newcommand{\LRS}[0]{\ensuremath{\SU{3}\otimes\SU{2}\otimes\SU{2}}}
\newcommand{\RSM}[0]{\ensuremath{\SU{3}\otimes\SU{2}}}
\newcommand{\SM}[0]{\ensuremath{\SU{3}\otimes\SU{2}\otimes\U{1}}}

\newcommand{\FSUfive}[0]{\ensuremath{\mathcal{F}\SU{5}}}
\newcommand{\G}[1]{\ensuremath{G_{#1}}}

\newcommand{\SUN}[0]{\ensuremath{\SU{N>5}}}
\newcommand{\SON}[0]{\ensuremath{\SO{N>10}}}

\newcommand{\N}{\ensuremath{\mathcal{N}}}
\newcommand{\half}{\ensuremath{\frac{1}{2}}}

\newtheorem*{definition}{Definition}

\makeatletter
    \def\imod#1{\allowbreak\mkern5mu({\operator@font mod}\,\,#1)}
\makeatother

\begin{document}

\title{Systematic Investigations of the Free Fermionic Heterotic String Gauge
    Group Statistics: Layer 1 Results}

\author{D. Moore}
\email[~~]{douglas\_moore1@baylor.edu}
\affiliation{Department of Physics, Baylor University,\\ Waco, TX 76798, USA}

\author{J. Greenwald}
\email[~~]{jared\_greenwald@baylor.edu}
\affiliation{Department of Physics, Baylor University,\\ Waco, TX 76798, USA}

\author{T. Renner}
\email[~~]{timothy\_renner@baylor.edu}
\affiliation{Department of Physics, Baylor University,\\ Waco, TX 76798, USA}

\author{M. Robinson}
\email[~~]{matt.robinson.string@gmail.com}
\affiliation{Department of Physics, Baylor University,\\ Waco, TX 76798, USA}

\author{C. Buescher}
\email[~~]{cameron\_buescher@baylor.edu}
\affiliation{Department of Physics, Baylor University,\\ Waco, TX 76798, USA}

\author{M. Janas}
\email[~]{janas@augsburg.edu}
\affiliation{Department of Physics, Augsburg College,\\ Minneapolis, MN 55454,
    USA}
\affiliation{Department of Physics, Baylor University,\\ Waco, TX 76798, USA}

\author{G. Miller}
\email[~]{gunnarmiller@gmail.com}
\affiliation{Department of Physics, University of Idaho,\\ Boise, ID 83702, USA}
\affiliation{Department of Physics, Baylor University,\\ Waco, TX 76798, USA}

\author{S. Ruhnau}
\email[~]{scott\_ruhnau@baylor.edu}
\affiliation{Department of Physics, Baylor University,\\ Waco, TX 76798, USA}

\author{G. Cleaver}
\email[~]{gerald\_cleaver@baylor.edu}   
\affiliation{Department of Physics, Baylor University,\\ Waco, TX 76798, USA}

\pacs{}
\preprint{BU-HEPP-11-05}
\preprint{CASPER-11-07}

\begin{abstract}
Using software under development at Baylor University, we explicitly construct
all layer 1 gauge, weakly coupled free fermionic heterotic string models up to
order $22$ in four large space-time dimensions. The gauge models consist
primarily of gauge content making a systematic construction process efficient.
We present an overview of the model building procedure, redundancies in the
process, methods used to reduce such redundancies and statistics regarding the
occurrence of various combinations of gauge group factors and GUT groups.
Statistics for both $\N=4$ and $\N=0$ models are presented.
\end{abstract}

\maketitle
\section{Introduction}\label{sec:1}
Recent work puts the number of possible string derived models on the order of
$10^{500}$ \cite{Bousso:2000, Ashok:2003}. Consequently, any efforts to explore
this landscape of string vacua require the use of high-performance computing and
a choice of construction method. Each construction method has access to
different, overlapping regimes of the landscape; here we will focus on the
weakly coupled free fermionic heterotic string (WCFFHS) construction formalism
\cite{Antoniadis:1986, Antoniadis:1987, Kawai:1986_2}. The WCFFHS formalism has
produced some of the most phenomenologically viable models to date
\cite{Cleaver:1999, Lopez:1992, Faraggi:1989, Faraggi:1992, Antoniadis:1990,
Leontaris:1999, Faraggi:1991, Faraggi:1992_2, Faraggi:1992_3, Faraggi:1991_2,
Faraggi:1991_3, Faraggi:1995, Faraggi:1996, Cleaver:1997, Cleaver:1997_2,
Cleaver:1997_3, Cleaver:1998, Cleaver:1998_2, Cleaver:1998_3, Cleaver:1998_4,
Cleaver:1998_5, Cleaver:1999_2, Cleaver:1999_3, Cleaver:1999_4, Cleaver:2000,
Cleaver:2000_2, Cleaver:2001, Cleaver:2001_2, Cleaver:2002, Cleaver:2002_2,
Cleaver:2002_3, Perkins:2003, Perkins:2005, Cleaver:2008, Greenwald:2009,
Cleaver:2011} and is ideal for computer construction. Random examinations of the
landscape, using this formalism, have been performed in the past
\cite{Dienes:2006, Dienes:2007_2}; however, due to the many-to-one nature of
this construction a random survey of the input parameters has many endemic
problems that are non-trivial to address \cite{Dienes:2007}.  One way to deal
with these problems is to \textit{systematically} survey the valid input
parameters.

Two software frameworks, currently under development at Baylor University, are
being designed and used specifically for the purpose of performing such
systematic surveys of the WCFFHS landscape. One such framework, the Gauge
Framework, focuses on systematically building gauge models in four large
space-time dimensions. A detailed explanation of what is meant by a ``gauge
model'' is provided in \autoref{sec:2}. These surveys serve multiple purposes
including aiding in attempts at understanding and reducing the redundancies
inherent to the construction method. Furthermore we can use the results of these
searches to guide slower, more detailed surveys.

The results presented here are intended more as a proof-of-concept than as an
exhaustive analysis of the data. The framework is a work in progress but
currently generates approximately $400$ models per second. Future iterations are
projected to exceed $1000$ models per second and will have the capability of
generating models in any number of large spacetime dimensions, of any order and
at any layer.

A detailed account of the WCFFHS formalism is provided here from the perspective
of gauge model building. We include discussion of the input space and the
systematic generation of the inputs. To discuss redundancy we need to first
define what it means for a model to be ``unique"; this definition is provided
in \autoref{sec:2.3}. As mentioned, the many-to-one nature of this formalism
remains a difficult problem to overcome. We discuss several steps taken that
have reduced redundancy to a level that admits systematic surveys of gauge
models in \autoref{sec:2.4}. We conclude with statistics from a layer 1 survey
of WCFFHS models from order 2 through order 22. These statistics divide nicely
into two categories, correlations between the input space and the model space,
and the statistics of various properties of the low energy effective field
theories themselves independent of input. We look at the number of models
generated at each order, \autoref{sec:3.1}, as well as the number of group
factors of each rank in the unique models built, \autoref{sec:3.2}, and the
combinations of group factors and GUT groups, \autoref{sec:3.3}, that occur in
both SUSY and non-SUSY models. Many of the results are presented in a truncated
form for brevity. The full range of statistics and complete data sets will be
made available at
\href{http://homepages.baylor.edu/eucos/svp}{http://homepages.baylor.edu/eucos/svp}.

\section{Gauge Model Building}\label{sec:2}
The Gauge Framework focuses on the construction of gauge models. Further
discussion requires a more concrete definition of what a ``gauge model'' is.
\begin{center}
    \parbox{.8\textwidth}{
        \begin{definition}[Gauge Model]
            A model is a \textbf{gauge model} if it can be built from a set of
            basis vectors in which every basis vector beyond the all-periodic
            and SUSY basis vectors is bosonic, that is of the form
            $(\vec{0}^{10}~||~\vec{\alpha})$, within the free fermionic
            construction \cite{Antoniadis:1986, Antoniadis:1987, Kawai:1986_2,
            Kawai:1987}
        \end{definition}
    }
\end{center}

These models are in many ways some of the most simple models that one can build.
We can think of them as the basis from which more complex models can be built.
This makes them interesting as a starting point for systematic surveys. We
can use what we learn about these models to guide further searches. Here we
review the WCFFHS construction method with gauge models in mind.

Within the free fermionic framework two inputs are required, the set of basis
vectors, $\bm{A}$, and the GSO projection coefficient matrix, $\bm{k}$. In order
to systematically build these models we need to systematically build the input
set $\{\bm{A},\bm{k}\}$ ensuring that all of the modular invariance constraints
are met.

\subsection{Basis Vectors}\label{sec:2.1}
In 4 large spacetime dimensions, the basis vector set is defined as
\begin{equation}\label{eqn:1}
    \bm{A} = \left\{\vec{\alpha}_i \left|\right. \vec{\alpha}_i \in
        \mathbb{Q}^{32} \cap \left(-1,1\right]^{32} \right\},
\end{equation}
where $i = 1,2,\dots,l$\footnote{We do not allow for chiral Ising models here.
So all left- or right-moving real fermions can be paired to form left- or
right-moving complex fermions.}. For our purposes we will always take $l\ge3$ and
will refer to $L = l-2$ as the layer. Each of these basis vectors represents the
boundary conditions of complex worldsheet fermion degrees of freedom. We will be
taking $\alpha^j_i$ with $j = 1,\dots,10$ to represent the boundary conditions
of the left-moving supersymmetric string and $\alpha^j_i$ with $j = 11,..,22$ to
represent the right-moving bosonic string boundary conditions. The order of each
basis vector, $N_i$, is the smallest integer such that
\begin{equation}\label{eqn:2}
    N_i ~ \alpha^j_i = 0 \pmod{2}.
\end{equation}
Of course, the choices of these basis vectors are constrained by modular
invariance in such a way that
\begin{equation}\label{eqn:3}
    N_i ~ \vec{\alpha}_i^2 =
    \begin{cases}
        0 \imod{8} & \textrm{if } N_i \textrm{ even}\\
        0 \imod{4} & \textrm{if } N_i \textrm{ odd}
    \end{cases}
\end{equation}
and
\begin{equation}\label{eqn:4}
    N_{ij} ~ \vec{\alpha}_i \cdot \vec{\alpha}_j = 0 \imod{4},
\end{equation}
where $N_{ij}\equiv\textrm{LCM}(N_i,N_j)$.

Since we are dealing with $L=1$, we have three basis vectors, two of which will
always be the same for every basis vector set we generate:
\begin{itemize}
    \item The first basis vector, denoted $\mathds{1}$, is the all periodic
        boundary conditions:
        $(\vec{1}^{10} ~||~ \vec{1}^{22})$.
    \item The second basis vector is the SUSY generator, $\bm{S}$, which is
        $(1~(100)^3 ~||~ \vec{0}^{22})$.
\end{itemize}

Generating these basis vectors can introduce excessive redundancy hindering
systematic search algorithms. However, an algorithm for efficiently generating
only the modular invariant basis vectors was developed and implemented
\cite{Robinson:2008}. Further discussion of the redundancies of this process is
found in \autoref{sec:2.4}.

\subsection{GSO Projection}\label{sec:2.2}
Once a modular invariant set of basis vectors has been generated, the states can
be built. To do so we begin by generating all sectors as linear combinations of
the basis vectors with $m^i_j \in \mathbb{N}$ and $m^i_j < N_j$, namely
\begin{equation}\label{eqn:5}
    \vec{V}^j = \sum_i m^j_i\vec{\alpha}_i.
\end{equation}

The sectors describe how the worldsheet fermions, $f_j$, transform around
non-contractible loops on the worldsheet
\begin{equation}\label{eqn:6}
    f_j \longrightarrow  - \mathit{e}^{\mathit{i}\pi V^i_j} f_j.
\end{equation}

To each of these sectors we apply fermion number operators, $\vec{F}^i$,
\begin{equation}\label{eqn:7}
    \vec{Q}^i = \frac{1}{2} \vec{V}^i + \vec{F}^i
\end{equation}
to build the charges. We can then express the masses of the states in terms of
the charges as
\begin{subequations}
    \begin{equation}
        \alpha' m_{left}^2 = \frac{1}{2}\left(\vec{Q}^i_{left}\right)^2
            - \frac{1}{2}
    \end{equation}
    \begin{equation}
        \alpha' m_{right}^2 = \frac{1}{2}\left(\vec{Q}^i_{right}\right)^2 - 1
    \end{equation}
\end{subequations}
Because we are working at the string scale and only interested in the low-energy
effective theory, these states must be massless. Thus,
\begin{subequations}
    \begin{equation}
        \left(\vec{Q}^i_{left}\right)^2 = 1
    \end{equation}
    \begin{equation}
        \left(\vec{Q}^i_{right}\right)^2 = 2
    \end{equation}
\end{subequations}

However, once the states are constructed we must ensure that they are, in fact,
physical. This requires the application of a GSO projection, and hence the
specification of GSO projection matrix, $\bm{k}$. This matrix is, in our case,
$(L+2) \times (L+2)$ and is constrained by modular invariance:
\begin{subequations}\label{eqn:10}
    \begin{equation}\label{eqn:10a}
        k_{ij} + k_{ji} = \frac{1}{2}\vec{\alpha}_i\cdot\vec{\alpha}_j \imod{2}
    \end{equation}
and
    \begin{equation}\label{eqn:10b}
        k_{ii} + k_{i0} = \frac{1}{4}\vec{\alpha}_i\cdot\vec{\alpha}_i - s_i
            \imod{2},
    \end{equation}
\end{subequations}
with
\begin{equation}\label{eqn:11}
    N_jk_{ij} = 0 \imod{2}.
\end{equation}
It is clear from \autoref{eqn:10} that, in general, we have
$\frac{1}{2}(L+1)(L+2) + 1$ degrees of freedom in our choice of $\bm{k}$.
However, one of the degrees of freedom, our choice of the $k_{00}$ element, has
no effect on the model generated so we fix it to $1$. This reduces us back to
$\frac{1}{2}(L+1)(L+2)$, meaning we can specify the lower-triangle of our GSO
projection matrix. There is however, a caveat; not every choice of the
lower-triangle yields a modular invariant matrix. In layer 1 models
specifically, there are eight possible lower-triangles each of which specifies
a GSO projection matrix. However, when the additional basis vector is odd, only
two are modular invariant. We can represent this freedom in $\bm{k}$ as
\begin{equation*}
    \left(
    \begin{array}{c|ccc}
                        & \mathds{1}    & \bm{S}    & \vec{\alpha}  \\\hline
        \mathds{1}      & 0             & 0         & 0             \\
        \bm{S}          & 2             & 0         & 0             \\
        \vec{\alpha}    & 2             & 2         & 0
    \end{array}
    \right)
\end{equation*}
where each entry represents the number of choices for that element of $\bm{k}$.
Now, when we consider an $\vec{\alpha}$ of odd order, and using
\autoref{eqn:11}, we see that the only choices of $k_{20}$ and $k_{21}$ that admit
$k_{i2} \in 2\mathbb{Z}$ values can be made. It is easy to see that, regardless of the
choice of $k_{21}$, $k_{12} = k_{21}$, because the forms of $\bm{S}$ and
$\vec{\alpha}$ presupposed. Consequently, because $k_{21} \in
\left\{0,1\right\}$ only $k_{21}=0$ is admitted. This takes the number of
admissible $\bm{k}$ choices to four. We can apply a similar analysis to the
$k_{20}$ choice. We know that $k_{20} \in \left\{0,1\right\}$, and that this
choice determines both $k_{02}$ and $k_{22}$. It has been shown that $k_{02}$
and $k_{22}$ are either both even or both odd \cite{Renner:2011}.  It is easy to
see that switching between choices of the $k_{20}$ flips the parity of $k_{02}$:
\begin{eqnarray}
    k_{ij} + k_{ji} =& \beta \imod{2}\\
    k'_{ij} + k'_{ji} =& \beta \imod{2}
\end{eqnarray}
where $k'_{ij} = k_{ij} + 1 \imod{2}$. Taking the difference between these
two, we get
\begin{eqnarray}
    k'_{ij} - k_{ij} =& ~k_{ji} - k'_{ji} \imod{2}\\
    k'_{ji} =& k_{ji} - 1 \imod{2}.
\end{eqnarray}
Thus, either the choice of $k_{20} = 0$ or $k_{20} = 1$ gives us even $k_{i2}$.
So, this again halves the number of accessible GSO projection matrices to two.

We then simply apply the GSO projection,
\begin{equation}
    \vec{\alpha}_i \cdot \vec{Q}^j = \sum_{l = 1}^{L+2} m^j_l k_{il} + s_i
        \imod{2}
\end{equation}
with $s_i$ being the space-time component of $\vec{\alpha}_i$. Now we are in
a position to consider the space-time supersymmetry of these models.

From a model building perspective, the simplest way to determine the number of
supersymmetries in a model is to count the number of gravitinos in the model.
Gravitinos have a specific form,
\begin{equation*}
    \begin{array}{c}
        \left(
            \frac{1}{2}
            \left(\pm\frac{1}{2} 0  0 \right)
            \left(\pm\frac{1}{2} 0  0 \right)
            \left(\pm\frac{1}{2} 0  0 \right)
            ~||~\vec{0}^{22}
        \right)
    \end{array}
\end{equation*}
when considering $SU(2)^6$ worldsheet SUSY. In our models, these states can only
arise from the SUSY sector generated by $\bm{S}$. Half of these eight gravitinos
have positive inner products with the SUSY generator, so when the GSO projection
is carried out, exactly half of the gravitinos are ejected from the model; the
half that is kept is determined by the $k_{22}$ element of $\bm{k}$.  Now,
because the additional basis vector, $\vec{\alpha}$ has a zero inner product
with the charges, the gravitino states are only kept if $k_{21}$ zero. Thus, in
these models, for a given set of basis vectors, if $k_{2i} = 0$ for any $i>2$,
SUSY is preserved at $\N = 4$ and is broken to $\N = 0$ otherwise.

We saw previously that if the $\vec{\alpha}_i$ is of odd order, $k_{2i}$ is
necessarily 0, thus at layer 1 there are no odd order, $\N=0$ models.

\subsection{Uniqueness}\label{sec:2.3}
When considering uniqueness of models, both gauge and matter content should be
considered. The nice thing about gauge models is that, when the model is
supersymmetric, models with the same gauge group will always have the same
matter spectrum and are thus identical. This, however, is not true for non-SUSY
models so, in general, to consider uniqueness we must investigate the matter
content of these models. Fortunately, the exact particle spectrum of these
models is not of interest here. We are only concerned with the gauge content and
whether the model is supersymmetric. From that, we can use additional software
to prepend left-movers to our basis vectors and build models using these gauge
models as a starting point. When there is no left-right pairing, the new models
will either keep or break the gauge group of their base gauge models. So, for
our purposes we will define uniqueness as follows:
\begin{center}
    \parbox{.8\textwidth}{
        \begin{definition}[Uniqueness]
            A model is considered unique if no other model has been previously
            generated with both the same gauge group and number of space-time
            supersymmetries.
        \end{definition}
    }
\end{center}
By this we mean that as we generate models, if we build one that has
a combination of gauge groups and number of space-time supersymmetries that has
not yet been created we retain it. However, any model after that with the same
combination of gauge states and SUSY will be discarded. This has an impact on
the statistics of the non-SUSY models which will be discussed in more detail in
subsequent sections.

We can easily determine the maximum number of unique models that can be built by
considering that only simply-laced gauge groups can appear and there are no rank
cuts \cite{Robinson:2008}, thus the total rank must be $22$ in $D=4$.
Determining all of the combinations of simply-laced gauge groups whose rank sums
to $22$ and doubling that, for SUSY and non-SUSY, gives us at most $48952$
unique gauge models. Of course, it is unlikely that all of these combinations
can exist, especially at $L=1$. In fact, if we perform the same analysis on
$D=10$ we find that $5714$ models could occur, but it is well known that only
$9$ models are realized by the $D=10$ heterotic landscape \cite{Kawai:1986},
even when considering full matter content. However, because the $D=4$
landscape is much more complex, we should expect a higher occurrence of
unique models than at $D=10$. These calculations have been performed for
$D=4$ through $D=10$ and are provided in \autoref{table:1}.
\setlength{\extrarowheight}{.5ex}
\begin{longtable}{|c|c|}
    \caption{Maximum Number of Unique Simply-Laced Gauge Models in $D$ Large
        Space-time Dimensions
    }\label{table:1}\\

    \hline
    $\bm{D}$ & \textbf{\# of Models} \\
    \hline\hline

    \endfirsthead
    \endhead
    \multicolumn{2}{|r|}{\textbf{Continued }$\longrightarrow$}\\\hline
    \endfoot
    \endlastfoot
    10 & 5714  \\\hline
     9 & 4140  \\\hline
     8 & 11988 \\\hline
     7 & 8576  \\\hline
     6 & 24508 \\\hline
     5 & 17341 \\\hline
     4 & 48952 \\\hline
\end{longtable}

\subsection{Redundancies}\label{sec:2.4}
The free fermionic construction formalism has the inherent problem of redundancy;
the mapping from input space to output space is many-to-one. This property is
what condemns random surveys and remains a problem for systematic searches.
Reducing these redundancies will bring systematic surveys within current
technological limits. Both the basis vectors and the GSO projection coefficients
present redundancies that can be accounted for and removed in many cases.

The systematic generation of basis vectors admits redundancy in at least two
ways, permutations and charge conjugacy. Permutations of the elements of a basis
vector leave the mapping invariant as long as the same permutation is applied to
each of the basis vectors in the set, i.e.
\begin{equation*}
    \left\{
        \begin{array}{c c c c c c c c c c c}
            ( & \vec{0}^{10} & || & 1 & 1 & 1 & 1 & 0 & 0 & \vec{0}^{18} & )\\
            ( & \vec{0}^{10} & || & 1 & 1 & 0 & 0 & 1 & 1 & \vec{0}^{18} & )\\
        \end{array}
    \right\}
    \cong
    \left\{
        \begin{array}{c c c c c c c c c c c}
            ( & \vec{0}^{10} & || & 1 & 1 & 0 & 1 & 0 & 1 & \vec{0}^{18} & )\\
            ( & \vec{0}^{10} & || & 1 & 1 & 1 & 0 & 1 & 0 & \vec{0}^{18} & )\\
        \end{array}
    \right\}.
\end{equation*}
Here the third column of the right-movers was switched with the sixth. These two
sets will generate the same output, given that the same GSO projection matrix is
chosen. A scheme for removing these permutation redundancies was developed in
\cite{Robinson:2008}. Additionally, we can always flip the signs of the charges
as long as that change does not remove modular invariance, i.e.
\begin{equation*}
    \left\{
        \begin{array}{c c c c c c c c c c c c c}
            ( & \vec{0}^{10} & || & 1 & 1 & 1 & 1 & 0 & 0 & 0 & 0 & \vec{0}^{18}
              & )\\

            ( & \vec{0}^{10} & || & \frac{2}{3} & \frac{2}{3} & -\frac{2}{3}
              & 0 & 0 & 0 & 0 & 0 & \vec{0}^{18} & )\\

            ( & \vec{0}^{10} & || & \half & \half & \half & -\half & \half
              & \half & -\half & -\half &  \vec{0}^{18} & )\\
        \end{array}
    \right\}
    \cong
    \left\{
        \begin{array}{c c c c c c c c c c c c c}
            ( & \vec{0}^{10} & || & 1 & 1 & 1 & 1 & 0 & 0 & 0 & 0 & \vec{0}^{18}
              & )\\

            ( & \vec{0}^{10} & || & \frac{2}{3} & \frac{2}{3} & \frac{2}{3}
              & 0 & 0 & 0 & 0 & 0 & \vec{0}^{18} & )\\

            ( & \vec{0}^{10} & || & \half & \half & -\half & -\half & \half
              & \half & \half & \half &  \vec{0}^{18} & )\\
        \end{array}
    \right\}
\end{equation*}
This is referred to as charge conjugacy and does not change the gauge group.
These two redundancies are not sufficient to completely remove the many-to-one
nature of the mapping, and studies are currently under way to find more sources
of basis vector redundancy.

Systematically generating the GSO projection matrices also introduces two
redundancies. The first redundancy is in our choice of which $4$ gravitinos to
remove via the $\bm{S}$ GSO projection. The choice does not affect the number of
supersymmetries nor what gauge group the model possesses and is specified by our
choice of $k_{10}$. By our definition of uniqueness, simply changing the
$k_{10}$ value will leave the model the same. Recall that $k_{10} \in \{0,1\}$,
so we can choose to only build the $\bm{k}$'s with $k_{10} = 0$. This means we
now only have $4$ choices of $\bm{k}$ for layer 1, even order basis vector sets
and $1$ for odd order sets. The second redundancy is in our choice of $k_{20}$
for layer 1, even order sets. Models built with either choice of $k_{20}$ are
identical by our definition. Why this is the case has not been shown
analytically, but has been confirmed up to order 22. This reduces our choices
down to $2$ and $1$ for layer 1, even and odd basis vector sets, respectively.

The result of accounting for these redundancies is a significant improvement in
the volume of models that must be built. \autoref{table:2} depicts the effects
of these redundancies. One thing to note is that each of these affects even and
odd orders to differing extents. However, if we account for all of them the
result is that the number of models that must be built (not the number of unique
models) at orders $2N$ and $2N+1$ are of the same order.

\begin{longtable}{|c|c|c|c|c|}
    \caption{\textbf{Number of $L=1$ Models} -
        For each order we list the most models possible and number of models
        after the permutation, charge conjugacy and GSO projection redundancies
        are accounted for. These are not necessarily distinct models, in fact
        the majority are still redundant.
    }\label{table:2}\\

    \hline
    \textbf{N}  & \textbf{Initial Models}   & \textbf{Permutation}
                & \textbf{Charge Conjugacy} & \textbf{GSO Projection}
                \\\hline\hline
    \endfirsthead

    \hline
    \textbf{N}  & \textbf{Initial Models}   & \textbf{Permutation}
                & \textbf{Charge Conjugacy} & \textbf{GSO Projection}
                \\\hline\hline
    \endhead

    \multicolumn{5}{|r|}{\textbf{Continued }$\longrightarrow$}\\\hline
    \endfoot
    \endlastfoot

     2 & $8.39\times10^6$    & $20$         & $20$      & $10$      \\\hline
     3 & $3.14\times10^{10}$ & $47$         & $7$       & $7$       \\\hline
     4 & $3.76\times10^{13}$ & $640$        & $152$     & $76$      \\\hline
     5 & $2.38\times10^{15}$ & $873$        & $55$      & $55$      \\\hline
     6 & $2.63\times10^{17}$ & $8292$       & $772$     & $386$     \\\hline
     7 & $3.91\times10^{18}$ & $9352$       & $328$     & $328$     \\\hline
     8 & $1.48\times10^{20}$ & $71724$      & $3748$    & $1874$    \\\hline
     9 & $9.85\times10^{20}$ & $70759$      & $1679$    & $1679$    \\\hline
    10 & $2.00\times10^{22}$ & $463948$     & $16172$   & $8086$    \\\hline
    11 & $8.14\times10^{22}$ & $413948$     & $7339$    & $7339$    \\\hline
    12 & $1.10\times10^{24}$ & $2434404$    & $62704$   & $31352$   \\\hline
    13 & $3.21\times10^{24}$ & $2007773$    & $28979$   & $28979$   \\\hline
    14 & $3.28\times10^{25}$ & $10756336$   & $223020$  & $111510$  \\\hline
    15 & $7.49\times10^{25}$ & $8378335$    & $104453$  & $104453$  \\\hline
    16 & $6.19\times10^{26}$ & $41719604$   & $730020$  & $365010$  \\\hline
\end{longtable}

\section{Statistics}\label{sec:3}
Traditionally, the collection of string derived, low energy effective field
theories (LEEFTs) is referred to as the landscape. However, because we are
interested less in the field theories and more in the mapping from the WCFFHS
input space to this landscape, we can consider only those LEEFTs that are mapped
to by a particular input sub-space; namely the layer 1, order 2 through 22 gauge
input space. For our purposes it is sufficient to refer to the LEEFTs that these
inputs map to as the ``layer 1 landscape." We can then look at the relationships
between the input and output spaces as well as the mapping between them.

Using the WCFFHS formalism, we constructed all unique, layer 1 gauge models from
order 2 through 22. This amounted to $68$ SUSY and $502$ non-SUSY models and
required a total of $31,863,121$ models to be built. Of all of the group
combinations found, $50$ had both SUSY and non-SUSY realizations. In this
section we review the statistics for these $570$ models as well as how we may
use these results to improve further surveys and what LEEFTs are
accessible from these types of inputs.

\subsection{Model Generation and Redundancy}\label{sec:3.1}
Here we look at relationships between model generation, the basis vector
order and redundancy. Strictly speaking, these relationships have no physical
meaning; however, they are important when creating algorithms for the systematic
generation of WCFFHS models, particularly for studies into how redundancies
manifest themselves in the gauge input space.

In our model building process, all models of a particular order are generated
before progressing to the next. This allows us to ask how the number of unique
models generated is dependent on the order. There is a subtlety to these
questions in that any model can be, in general, generated at other orders.
However, because we have imposed an ordering on the build process this
inherently gives preference to lower orders. This has the advantage of improving
the efficiency of the build process and does not affect statistics beyond the
physically meaningless question of ``at what order was this model generated?"

We know from \autoref{sec:2.2} that there are no odd-order $\N = 0$
models. This immediately suggests that there is a difference in the way SUSY and
non-SUSY models are generated at each order. This difference can be seen in
\autoref{figure:1} where the number of unique models generated is plotted with
respect to order for both SUSY and non-SUSY data sets. We see that a statistical
majority of SUSY models are generated at low order, from order 2 through 12,
while a statistically significant number of non-SUSY models are generated
through 22.

\begin{center}
    \captionof{figure}{\textbf{Number of New Models at Each Order} - Order
        6 generates the most unique SUSY models at 18, and order 12 generates
        the most non-SUSY with 96. Note that the non-SUSY curve only has data
        for even orders because no odd order non-SUSY models exist.
    }\label{figure:1}
    \includegraphics[width=3in]{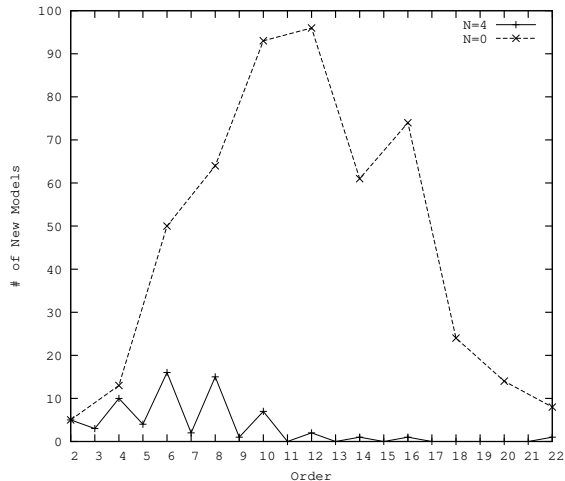}
\end{center}

One may also be interested in how higher orders subsume lower orders. That is,
because higher orders admit a significantly higher number of potential models,
one might suspect that higher orders may well contain all of the models
generated at lower orders. To verify this we look at the number of unique models
generated at each order as well as the number of models that were generated at
lower orders but are absent from higher orders, \autoref{figure:2}.

\begin{center}
    \captionof{figure}{\textbf{Number of Additional and Absent Models at Each
        Order} - At each order we look at the number of models generated in
        addition to the models previously created as well as the number of
        models that are absent at that order. Note that no $\N=0$ models are
        generated at odd orders so, for brevity, those orders are not plotted. 
    }\label{figure:2}
    \includegraphics[width=\textwidth]{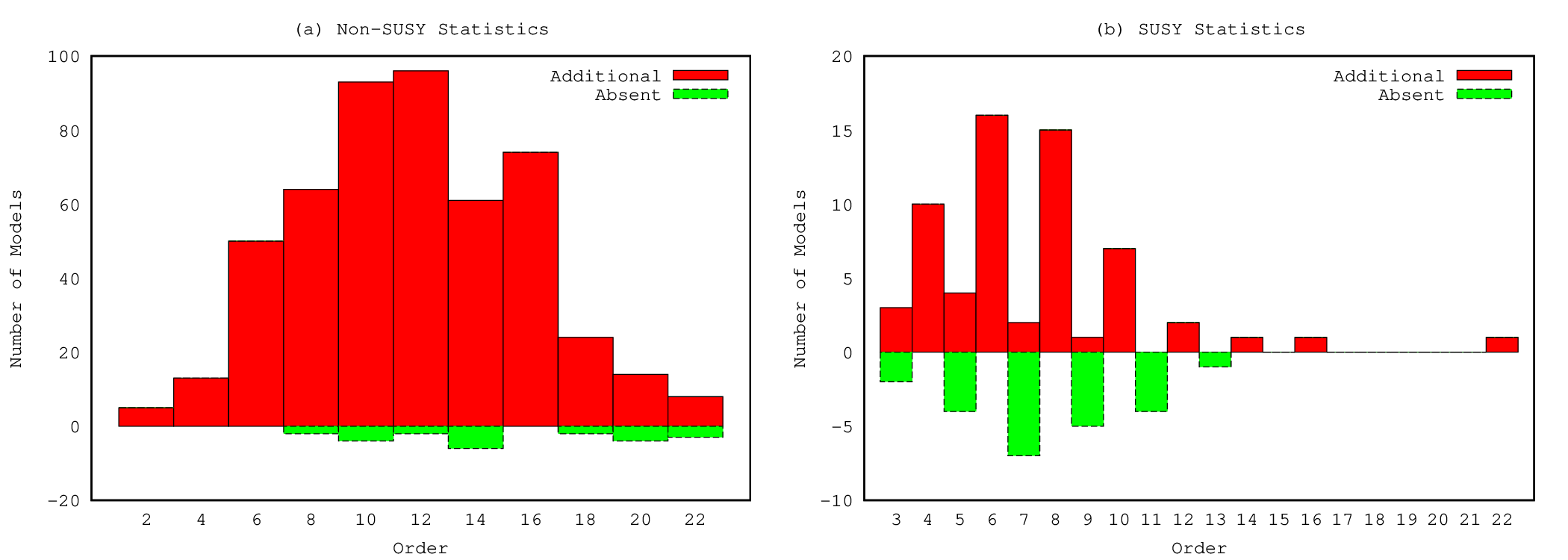}
\end{center}

This information can be used to more efficiently generate models. For example,
even order SUSY models completely subsume lower orders. This means that we could
simply build SUSY order 22 and we would get everything below it. That reduces
the number of SUSY models that must be built from approximately $1.82\times10^7$
to $8.4\times10^6$, roughly in half. This is not quite as nice for non-SUSY
models in that we would have to build orders $16$ through $22$. This amounts to
$98.89\%$ of the total number of models. Only SUSY would benefit from this
approach. Unfortunately, there is no known way to predict which orders subsume
lower orders.

\subsection{Group Distribution Statistics}\label{sec:3.2}
We now focus on specific properties of the LEEFTs, in particular how group
factors of each rank manifest themselves across the layer 1 landscape. We begin
by considering the number of models with a group factor of a particular
rank, $M_n$, for each, SUSY and non-SUSY, data set, \autoref{figure:3}.
\begin{center}
    \captionof{figure}{\textbf{Number of Models with Factors of Each Rank} - For
        each rank and class of gauge group, the number of models with at least
        one factor of that type is plotted.  The label on each bar is the total
        number of models with at least one group of that rank. The plots for the
        SUSY and Non-SUSY models are provided for comparison.
    }\label{figure:3}
    \includegraphics[width=7in]{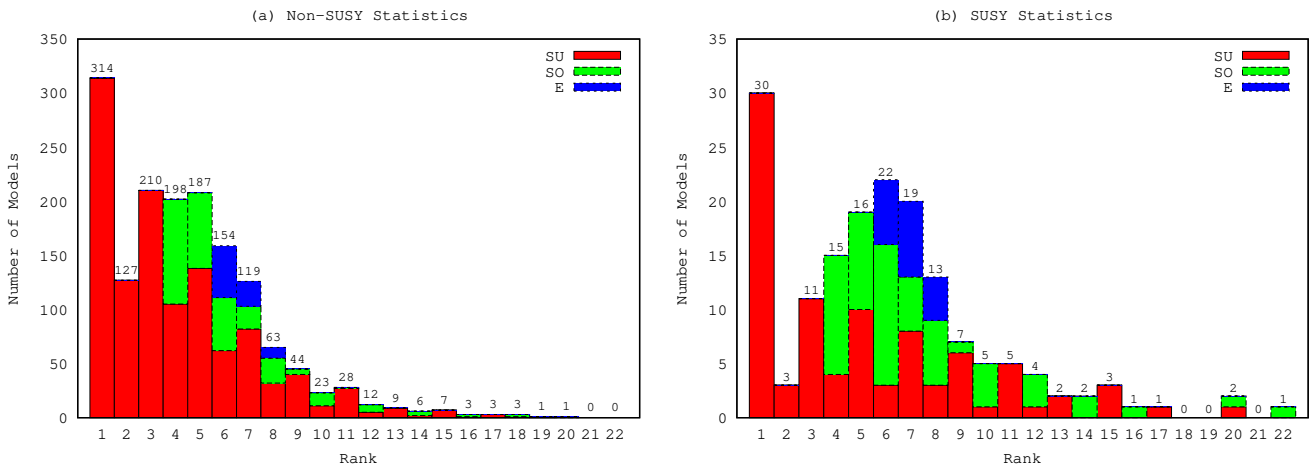}
\end{center}
\SU{2} is highly prevalent in both datasets because it is relatively simple to
generate. It amounts to a single, disjoint charge in our gauge states and
consequently occurs often when groups are broken.

For the $\N=0$ models we see that $M_n > M_{n+2}$ for all classes of group
factor, \SU{N}, \SO{N} and \E{N}. However, this trend only occurs for \SU{N} of
odd rank up to $n=11$. Additionally, we can see $M_{2n-1} > M_{2n}$ up to
$n=10$ for non-SUSY models.

This does not speak to how the factors are distributed amongst the models. Of
the non-SUSY models, 314 have at least one factor of \SU{2}, but generally we can
expect more than one for a particular model, approximately 1.74 on average. The
average number of factors of each rank, $\bar{M}_n$, is plotted in
\autoref{figure:4}.
\begin{center}
    \captionof{figure}{\textbf{Average Number of Factors of Non-Abelian Groups}
        - For each rank, the average number of factors for each class of groups
        is plotted for each set of statistics, (a) Non-SUSY Models and (b) SUSY
        Models.
    }\label{figure:4}
    \includegraphics[width=7in]{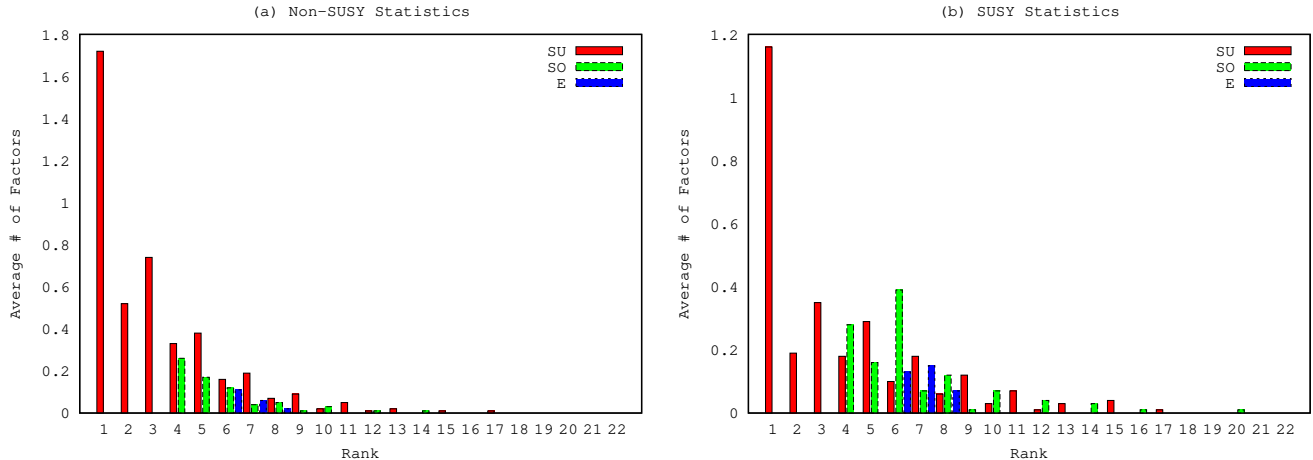}
\end{center}

\subsection{Group Combinations}\label{sec:3.3}
In this subsection we present statistics for occurrence of specific group
factors in various combinations across the layer 1 landscape. As well as
combinations of two group factors, we look at combinations of specific compound
factors in conjunction with single and other compound factors. Such compound
factors include $\EE{6}{6}$, $\G{PS}\equiv\PS$ (Pati-Salam), $\G{LRS}\equiv\LRS$
(Left-Right Symmetric), $\G{SM}\equiv\SM$ (Standard Model), and
$\G{RSM}\equiv\RSM$ (Reduced Standard Model).  We also include
$\FSUfive\equiv\FSU{5}$, though, because we are not considering matter content,
we can only say that the model has the $\FSUfive$ gauge group; it may not
actually be $\FSUfive$.

Recall that the modular invariance constraints and redundancies lead to
two GSO projection matrices for even order models and only one for odd order
models. In either case, the GSO projection that admits $\N=4$ SUSY is consistent,
as discussed in \autoref{sec:2.2}. The SUSY landscape exhibits 68 unique models.
From these, the percentage of models exhibiting each combination of group
factors at least once is calculated as a straight percentage of the 68 models.
These statistics are provided in \autoref{table:5}.

\begin{longtable}{|c|c|c|c|c|c|c|c|c|c|c||c|c|c|c|c|c|}
    \caption{\textbf{$\bm{\N=4}$ Gauge Group Combinations (All Models)} -
        The percentage of all unique $\bm{\N=4}$ models with each
        combination of gauge groups is tabulated. For example, 11.76\% of the 68
        unique SUSY models have the combination $\SU{4}\otimes\U{1}$ at least
        once.
    }\label{table:5}\\

    \hline
    $N=68$      & \U{1}     & \SU{2}    & \SU{3}    & \SU{4}    & \SU{5}
                & \SUN      & \SO{8}    & \SO{10}   & \SON      & \E{N}
                & \EE{6}{6} & \FSUfive  & \G{PS}    & \G{LRS}   & \G{RSM}
                & \G{SM}    \\\hline\hline
    \endfirsthead

    \hline
                & \U{1}     & \SU{2}    & \SU{3}    & \SU{4}    & \SU{5}
                & \SUN      & \SO{8}    & \SO{10}   & \SON      & \E{N}
                & \EE{6}{6} & \FSUfive  & \G{PS}    & \G{LRS}   & \G{RSM}
                & \G{SM}    \\\hline\hline
    \endhead
    \multicolumn{17}{|r|}{\textbf{Continued }$\longrightarrow$}\\\hline
    \endfoot
    \endlastfoot

    \U{1}       & 29.41 & 11.76 & 4.41  & 11.76 & 5.88  & 36.76 & 4.41  & 7.35  & 11.76 & 5.88  & 1.47  & 5.88  & 2.94  & 0     & 0     & 0     \\\hline
    \SU{2}      & --    & 29.41 & 0     & 7.35  & 0     & 32.35 & 11.76 & 7.35  & 25.00 & 10.29 & 0     & 0     & 1.47  & 0     & 0     & 0     \\\hline
    \SU{3}      & --    & --    & 2.94  & 0     & 1.47  & 2.94  & 0     & 0     & 0     & 0     & 0     & 1.47  & 0     & 0     & 0     & 0     \\\hline
    \SU{4}      & --    & --    & --    & 5.88  & 0     & 11.76 & 1.47  & 4.41  & 1.47  & 2.94  & 1.47  & 0     & 2.94  & 0     & 0     & 0     \\\hline
    \SU{5}      & --    & --    & --    & --    & 4.41  & 2.94  & 0     & 0     & 0     & 0     & 0     & 4.41  & 0     & 0     & 0     & 0     \\\hline
    \SUN        & --    & --    & --    & --    & --    & 33.82 & 7.35  & 12.24 & 16.18 & 8.82  & 1.47  & 2.94  & 4.41  & 0     & 0     & 0     \\\hline
    \SO{8}      & --    & --    & --    & --    & --    & --    & 4.41  & 0     & 5.88  & 1.47  & 0     & 0     & 0     & 0     & 0     & 0     \\\hline
    \SO{10}     & --    & --    & --    & --    & --    & --    & --    & 2.94  & 0     & 2.94  & 0     & 0     & 0     & 0     & 0     & 0     \\\hline
    \SON        & --    & --    & --    & --    & --    & --    & --    & --    & 17.65 & 13.24 & 0     & 0     & 0     & 0     & 0     & 0     \\\hline
    \E{N}       & --    & --    & --    & --    & --    & --    & --    & --    & --    & 11.76 & 1.47  & 0     & 1.47  & 0     & 0     & 0     \\\hline\hline
    \EE{6}{6}   & --    & --    & --    & --    & --    & --    & --    & --    & --    & --    & 0     & 0     & 0     & 0     & 0     & 0     \\\hline
    \FSUfive    & --    & --    & --    & --    & --    & --    & --    & --    & --    & --    & --    & 4.41  & 0     & 0     & 0     & 0     \\\hline
    \G{PS}      & --    & --    & --    & --    & --    & --    & --    & --    & --    & --    & --    & --    & 0     & 0     & 0     & 0     \\\hline
    \G{LRS}     & --    & --    & --    & --    & --    & --    & --    & --    & --    & --    & --    & --    & --    & 0     & 0     & 0     \\\hline
    \G{RSM}     & --    & --    & --    & --    & --    & --    & --    & --    & --    & --    & --    & --    & --    & --    & 0     & 0     \\\hline
    \G{SM}      & --    & --    & --    & --    & --    & --    & --    & --    & --    & --    & --    & --    & --    & --    & --    & 0     \\\hline\hline
    Total       & 45.58 & 44.12 & 4.41  & 16.18 & 5.88  & 54.41 & 16.18 & 13.24 & 44.12 & 23.53 & 2.94  & 5.88  & 5.88  & 0     & 0     & 0     \\\hline
\end{longtable}

We included the \G{LRS}, \G{RMS} and \G{SM} entries for completeness. While they
are identically zero for $\N=4$ models, this is not true for $\N=0$ models, thus
we include them for future consistency.

\SU{3} never occurs in tandem with \SU{2}. This means there is no Standard Model
gauge group in the SUSY layer 1 landscape, as defined here. Pati-Salam and
$\FSUfive$ occur in an equal number across the SUSY landscape but never in the
same model. The only compound factor that occurs more than once in any model is
$\FSUfive$ and it does so $75\%$ of the time, though this only amounts to $3$
models in total.

Turning our attentions to $\N=0$ models, we can perform the same statistical
analysis we did above. This time, however, we note that there are $502$ unique
non-SUSY models.

\begin{longtable}{|c|c|c|c|c|c|c|c|c|c|c||c|c|c|c|c|c|}
    \caption{\textbf{$\bm{\N=0}$ Gauge Group Combinations (All Models)} - The
        percentage of all unique $\N=0$ models with each combination of gauge
        groups is tabulated. Here each value is calculated against the $502$
        $\N=0$ models, i.e. $\SO{10}\otimes\SU{5}$ occurs in $1.00\%$ of these
        $502$ models.
    }\label{table:6}\\

    \hline
    $N=502$     & \U{1}     & \SU{2}    & \SU{3}    & \SU{4}    & \SU{5}
                & \SUN      & \SO{8}    & \SO{10}   & \SON      & \E{N}
                & \EE{6}{6} & \FSUfive  & \G{PS}    & \G{LRS}   & \G{RSM}
                & \G{SM}    \\\hline\hline
    \endfirsthead

    \hline
                & \U{1}     & \SU{2}    & \SU{3}    & \SU{4}    & \SU{5}
                & \SUN      & \SO{8}    & \SO{10}   & \SON      & \E{N}
                & \EE{6}{6} & \FSUfive  & \G{PS}    & \G{LRS}   & \G{RSM}
                & \G{SM}    \\\hline\hline
    \endhead
    \multicolumn{17}{|r|}{\textbf{Continued }$\longrightarrow$}\\\hline
    \endfoot
    \endlastfoot

    \U{1}       & 75.70 & 49.80 & 25.30 & 40.64 & 20.92 & 63.75 & 11.75 & 13.35 & 10.36 & 10.36 & 1.39  & 20.92 & 15.54 & 8.76  & 14.74 & 14.74 \\\hline
    \SU{2}      & --    & 42.63 & 14.74 & 24.50 & 12.35 & 39.24 & 12.55 & 8.17  & 12.75 & 9.96  & 1.00  & 12.35 & 9.36  & 4.78  & 8.76  & 8.76  \\\hline
    \SU{3}      & --    & --    & 13.94 & 11.95 & 10.76 & 15.74 & 1.00  & 1.20  & 0     & 0.60  & 0     & 10.76 & 3.78  & 6.57  & 8.96  & 8.96  \\\hline
    \SU{4}      & --    & --    & --    & 19.12 & 9.56  & 28.09 & 6.37  & 5.98  & 4.18  & 3.98  & 0.60  & 9.56  & 8.37  & 3.78  & 7.17  & 7.17  \\\hline
    \SU{5}      & --    & --    & --    & --    & 8.76  & 12.35 & 0.80  & 1.00  & 0     & 0.60  & 0     & 8.76  & 2.59  & 4.38  & 7.17  & 7.17  \\\hline
    \SUN        & --    & --    & --    & --    & --    & 31.27 & 9.76  & 10.16 & 9.56  & 7.97  & 0.40  & 12.35 & 9.36  & 3.00  & 7.17  & 7.17  \\\hline
    \SO{8}      & --    & --    & --    & --    & --    & --    & 4.58  & 1.79  & 4.98  & 3.78  & 0     & 0.80  & 2.39  & 0     & 0     & 0     \\\hline
    \SO{10}     & --    & --    & --    & --    & --    & --    & --    & 2.59  & 1.20  & 2.19  & 0.40  & 1.00  & 2.39  & 0     & 0     & 0     \\\hline
    \SON        & --    & --    & --    & --    & --    & --    & --    & --    & 5.98  & 5.38  & 0.20  & 0     & 0.40  & 0     & 0     & 0     \\\hline
    \E{N}       & --    & --    & --    & --    & --    & --    & --    & --    & --    & 3.78  & 0.20  & 0.60  & 1.59  & 0     & 0     & 0     \\\hline
    \EE{6}{6}   & --    & --    & --    & --    & --    & --    & --    & --    & --    & --    & 0     & 0     & 0.20  & 0     & 0     & 0     \\\hline\hline
    \FSUfive    & --    & --    & --    & --    & --    & --    & --    & --    & --    & --    & --    & 8.76  & 2.59  & 4.38  & 7.17  & 7.17  \\\hline
    \G{PS}      & --    & --    & --    & --    & --    & --    & --    & --    & --    & --    & --    & --    & 2.99  & 0.80  & 1.99  & 1.99  \\\hline
    \G{LRS}     & --    & --    & --    & --    & --    & --    & --    & --    & --    & --    & --    & --    & --    & 1.99  & 3.78  & 3.78  \\\hline
    \G{RSM}     & --    & --    & --    & --    & --    & --    & --    & --    & --    & --    & --    & --    & --    & --    & 6.57  & 6.57  \\\hline
    \G{SM}      & --    & --    & --    & --    & --    & --    & --    & --    & --    & --    & --    & --    & --    & --    & --    & 6.57  \\\hline\hline
    Total       & 83.67 & 62.55 & 25.30 & 41.83 & 20.92 & 66.53 & 19.32 & 13.94 & 21.51 & 15.54 & 1.39  & 20.92 & 16.33 & 8.76  & 14.74 & 14.74 \\\hline
\end{longtable}

It is interesting to note that the occurrence of $\N=0$ group combinations is not
simply an extension of the $\N=4$. That is, groups that are less common in
$\N=4$ models are not necessarily less common in $\N=0$. We also find that
$\RSM$ combinations never occur with $\SO{N}$ nor $\E{N}$.

\section{Conclusion}\label{sec:3}
We have shown the computational feasibility of systematically generating models
of this type by creating all possible layer 1 gauge models through order 22.
While these are by no means phenomenologically realistic models, some show
properties that suggest possible avenues of investigations. Furthermore, several
redundancies in the construction process have been demonstrated and will be
explored further with future work. Additionally, the development process has
provided tools that can be used for higher layer surveys as well. One such
survey, examining the contents of the Layer 2 landscape through order 22, is
currently underway. In future work, we hope to extend these models by prepending
left-movers to determine how well these surveys can be used to guide more
exhaustive and detailed surveys excluding left-right pairing.

\bibliography{main}

\end{document}